\documentclass[copyright,creativecommons]{eptcs}
\providecommand{\event}{FROM 2026}

\usepackage{amsmath,amssymb,stmaryrd,txfonts,color,xcolor}

\newcommand{\N}{\mathbb{N}}

\newcommand{\true}{\mathit{true}}
\newcommand{\false}{\mathit{false}}
\newcommand{\Set}{\mathit{Set}}

\newcommand{\itree}{\mathit{itree}}
\newcommand{\pure}{\mathit{pure}}
\newcommand{\impure}{\mathit{impure}}
\newcommand{\CPO}{\mathit{CPO}}
\newcommand{\bind}{\mathit{bind}}
\newcommand{\fbind}{\mathit{fbind}}
\newcommand{\ccase}{\mathsf{case}}
\newcommand{\oof}{\mathsf{of}}
\newcommand{\eend}{\mathsf{end}}
\newcommand{\llet}{\mathsf{let}}
\newcommand{\iin}{\mathit{in}}
\newcommand{\ret}{\mathit{ret}}
\newcommand{\mdo}[1]{\mathrm{do}\ #1}

\newcommand{\inl}{\mathsf{inl}}
\newcommand{\inr}{\mathsf{inr}}
\newcommand{\while}{\mathsf{while}}
\newcommand{\whilefuel}{\mathsf{while\_fuel}}
\newcommand{\nbzeroes}{\mathit{nb\_zeroes}}
\newcommand{\iter}{\mathit{iter}}
\newcommand{\iif}{\mathsf{if}}
\newcommand{\even}{\mathit{even}}
\newcommand{\tthen}{\mathsf{then}}
\newcommand{\eelse}{\mathsf{else}}
\newcommand{\ttt}{*}

\newcommand{\statet}{\mathit{StateT}}
\newcommand{\get}{\mathit{get}}
\newcommand{\geteff}{\mathbf{get}}
\newcommand{\sput}{\mathit{put}}
\newcommand{\sputeff}{\mathbf{put}}

\newcommand{\trigger}{\mathit{trigger}}
\newcommand{\lub}{\mathit{lub}}

\newcommand{\fy}{\mathit{fy}}
\newcommand{\fz}{\mathit{fz}}
\newcommand{\fold}{\mathit{fold}}
\newcommand{\ffold}{\mathit{ffold}}
\newcommand{\floor}[1]{\llfloor{#1}\rrfloor}
\newcommand{\node}{\mathit{node}}

\newenvironment{notation}{\noindent \textbf{Notation.}}{}
\newtheorem{definition}{Definition}
\newtheorem{remark}{Remark}
\newtheorem{lemma}{Lemma}
\newtheorem{theorem}{Theorem}
\newtheorem{example}{Example}

\title{%Escaping the Coinduction Trap: \\ 
Domain Theory Meets Interaction Trees in Rocq}
\author{David Nowak
\institute{Univ.\ Lille, CNRS, Centrale Lille, UMR 9189 CRIStAL, F-59000 Lille, France}
\and
Vlad Rusu
\institute{Inria, Univ.\ Lille, CNRS, Centrale Lille, UMR 9189 CRIStAL, F-59000 Lille, France}
}

\def\titlerunning{%Escaping the Coinduction Trap: 
Domain Theory Meets Interaction Trees in Rocq}
\def\authorrunning{David Nowak \& Vlad Rusu}

\begin{document}

\maketitle
%TODO mandatory: add short abstract of the document
\begin{abstract}
We present a domain-theoretical  formalization of interaction trees in the Rocq prover. Unlike existing formalizations, ours does not rely on Rocq's built-in coinduction. Hence, we avoid complications occurring in earlier works, such as artificially including silent steps to comply with Rocq's productivity checker, treating monad laws as weak bisimulations, and coinductive bisimulation reasoning.
We define an inductive program equivalence relation as the congruence closure of a base relation on primitive effects with respect to action sequencing and least upper bounds. This enables reasoning about possibly non-terminating programs by reducing their equivalences to equivalences of their terminating approximations, which are then proved by induction.
This relation is proved correct: provided the base relation is  correct, equivalent computations have equal denotations in any monad that faithfully implements the effects.
We illustrate the framework by showing the equivalence of two programs encoding the Syracuse sequence, whose termination is an open mathematical conjecture.
\end{abstract}

\section{Introduction}
\label{sec:introduction}

Modern programming languages incorporate effects such as mutable state, exceptions, nondeterminism, input/output, and continuations. Monads offer a uniform representation of effectful computations \cite{Moggi89,Moggi91a} and are included, among others, in the Haskell programming language \cite{Wadler92}.
%However, while monads neatly capture individual effects, it is not immediately clear how interacting effects can be obtaned by combining their respective monads.
%A practical approach, implemented in Haskell, uses monad transformers \cite{Moggi91b}, which extend an existing monad with an additional effect, producing a new monad.
%However, monad transformers do not exist for all effects, and their definitions appear to be ad hoc.

%A more systematic and principled  approach is based on \emph{coproducts}, which combine two structures in the most general way, preserving the behavior of each of them, without imposing  additional interactions.
%The result is a monad that supports both effects in a modular way.
%Unfortunately, coproducts are not guaranteed to exist for all monads, and can be difficult to construct even when they exist.
%A case where coproducts do exist is that of \emph{free monads}~\cite{Swierstra08}.
A \emph{free monad} of a given effect corresponds to a signature of operations and represents all terms that can be formed from these operations, without interpretation.
%\textcolor{blue}{
This  yields several advantages: first, it allows to decouple the definition of effectful programs from their execution, in contrast to conventional monads where the two concerns are interwoven. This enables multiple interpretations of the same program: for example, a computation expressed as a free monad can be executed using a standard interpreter, or tested with mock handlers, all without altering the program itself. Second, the signatures of operations exposed by a free monad allow for equational reasoning, independently of its implementation.
%}
%,
%and the coproduct of two free monads is simply the free monad on the sum of their generating functors.

While free monads can easily be defined  in Haskell, their standard inductive definition cannot be transferred to Rocq, as it violates \emph{strict positivity}, which, in turn, compromises logical consistency. To circumvent this limitation, one can adopt \textit{freer} monads~\cite{KI15}, which preserve positivity while expressing the same idea: computations constructed from abstract operations. Unlike free monads, freer monads require only the object-level mapping of a signature and not the full functorial action.
%As with free monads, coproducts of freer monads always exist and are straightforward to construct: the coproduct of two freer monads is simply the freer monad over the sum of their operation signatures. 

%One notable framework based on freer monads is FreeSpec \cite{Letan20,Letan18}, which offers a principled approach to the modular verification of \emph{finite} computation of effectful programs in Rocq. 

\emph{Interaction-tree monads}~\cite{itree19} extend freer monads with coinduction, thereby enabling the representa\-tion of possibly infinite effectful computations.
Interaction trees, as well as an extension that offers special support for nondeterminism~\cite{Chappe25}, have been formalized in Rocq~\cite{Koh19,Silver23}, using Rocq's builtin coinductive features. This results in  a powerful foundation for formally verified semantics and equational reasoning. However, there is a price to pay for relying on Rocq's coinduction. It requires the incorporation of  artificial \emph{silent steps} 
in interaction trees in order to comply with Rocq's strict productivity checker. The monad laws only hold up to a 
\emph{weak bisimulation}, and the equational reasoning is performed modulo the weak bisimulation in question; all of which has significant consequences in practice.

For example, consider two programs that each compute a natural number if they terminate. To prove that the programs are equivalent, the corresponding interaction trees are traversed using a Rocq corecursive function, which encodes the program results into a coinductive version of the naturals; the coinductive type is required by Rocq as return type of a corecursive function. Completing the proof of program equivalence additionally requires a proof by coinduction that the results are weakly bisimilar. 
This rather complicated process is the consequence of relying on Rocq's builtin coinduction.

In this paper we present a novel formalization of interaction trees in Rocq.
Our approach is grounded in \emph{domain theory}, which
enables us to encode coinductive types and possibly non-terminating computations without using Rocq's coinductive features. Continuing on the example of  two programs computing natural numbers: in our approach, when the two programs are equivalent, they just compute  equal results. 

\smallskip

We prove equivalence of possibly non-terminating programs with effects according to an equivalence relation defined as follows.
Starting from a base relation that specifies the primitive effects, we take its inductively defined closure with respect to both the monadic structure and the formation of least upper bounds. Closure under least upper bounds enables reasoning about the equivalence between two possibly non-terminating programs by first reducing it to equivalences between  terminating approximations of the respective programs, which is then proved by induction. 

The equivalence relation is proved to be correct, in the sense that equivalent computations have equal denotations, where denotations interpret computations with given effects in monads implementing the corresponding effects. 
The correctness is proved once and for all, under the assumption that the base relation is itself correct; the correctness of the base relation is proved once per effect family. 
Overall, the correctness of the equivalence relation is independent of the proofs of equivalence for specific programs. 
%which is established on a case-by-case basis.

%Moreover, our formalization supports coproducts and modular equational reasoning for programs that combine multiple effects: the equations defining each individual effect are seamlessly integrated through the coproduct structure of interaction trees. 

We illustrate %the expressiveness of 
our framework
%by proving the correctness of a possibly non-terminating, stateful probabilistic program, and
by establishing the equivalence between two programs encoding the Syracuse sequence, whose termination is
%depends on 
an open mathematical conjecture.

\paragraph*{Paper Overview}
Section~\ref{sec:background} contains background on domain theory, starting with basic material and continuing with a domain-theoretical construction from our earlier paper~\cite{jlamp24} that we use for encoding interaction trees. Section~\ref{sec:monads} introduces CPO-valued monads, structures where values can be partially defined and where the sequencing operation is a continuous function. Examples of CPO-valued monads are presented. 
Section~\ref{sec:itree}  defines interaction trees in a domain-theoretical setting,
 and shows how the resulting construction is organized as a CPO-valued monad.
% The sequencing operation (\emph{bind}), and a \emph{fold} function for traversing and interpreting the effects
% of interaction trees, are defined using domain-theoretical techniques. 
We also introduce our generic equivalence relation and state its correctness.
%built on a base relation (specific to each effect family), closed under generic operations such as sequencing and least-upper bounds. The correctness of the relation with respect to denotations of interaction trees monads is proved.
% Section~\ref{sec:coprod} constructs the coproduct of interaction-tree monads by combining their effect signatures and shows how their equational reasoning techniques can be combined as well. 
We illustrate these concepts on an
example in Section~\ref{sec:examples} and present our conclusions in Section~\ref{sec:concl}, together with additional related work and future work directions. 
For  readability and generality the paper does not use Rocq code, but standard mathematical notation.
A complete implementation  in the Rocq proof assistant is available at:
\href{https://github.com/hidden-author/itreedomain/}{\texttt{\small https://github.com/hidden-author/itreedomain/}}.

\section{Background} 
\label{sec:background}
Domain theory is the object of several  textbooks~\cite{domains_and_lambda_calculi,mathematical_theory_of_domains,winskel93}.  We also present here a specific construction used for our definition of interaction trees, adapted from our earlier paper~\cite{jlamp24}. A distinctive feature  of domain theory is that all orders within %
are \emph{definition} orders: $x \preceq y $ means that $x$ is at most as defined as $y$.

\subsection{Elements of Domain Theory}
\label{sec:domain}
  A \emph{pointed partial order} (PPO) is a triple  $(C,\preceq,\bot)$   where $(C,\preceq)$ is a partially-ordered set (poset) and 
   $\bot \in C$ is the least element. 
A PPO $(C,\preceq,\bot)$ is \emph{flat} when $\preceq$ restricted to $C \setminus \{\bot\}$ is equality. That is, $c \preceq c'$ iff
$c = \bot$ or $c = c'$.
A set $S \subseteq C$ is \emph{directed} if $S \not = \emptyset$ and for all $x,y \in S$ there exists $z \in S$
  such that $x,y \preceq z$. 

\smallskip

  The \emph{least upper bound} of a set $S$ is
  denoted by $\lub\; S$.  
  A \emph{complete partial order} (CPO) is a PPO
  $(C, \preceq,\bot)$  such that 
each directed set $S \subseteq C$ has a least upper bound. For example, flat PPOs are CPOs.

\smallskip

In a CPO $(C,\preceq,\bot)$, an element $c^\circ$ is \emph{compact} (or \emph{finite}) whenever for all directed  $S \subseteq C$,
if $c^\circ \preceq \lub \ S$ then there exists $c \in S$ such that $c^\circ  \preceq c$. For example, $\bot$ is compact.
We denote by $C^\circ$ the  set of compacts of $(C,\preceq,\bot)$ and by $\preceq^\circ$ the restriction of  $\preceq$
to $C^\circ$. We call $(C^\circ,\preceq^\circ,\bot)$ the \emph{PPO of compacts} of $(C,\preceq, \bot)$.

\smallskip

For $c \in C$ we define $C^\circ_{c} := \{c^\circ \in C^\circ  \mid  c^\circ \preceq c\}$. 
Compact elements are essential in \emph{algebraic}
CPOs (ACPOs);  CPOs with the additional property that for all $c \in C$, the set $C^\circ_{c}$
is directed and $c = \lub \ C^\circ_{c}$. 

\smallskip

Given two CPOs $(D,\precsim,\bot)$ and $(C,\preceq,\Bot)$, a function $f : D \to C$ is \emph{continuous} when
it is monotonic and \emph{commutes with $\lub$s}:
for all directed $S \subseteq D$, $f \ (\lub\ S) = \lub \ (f \ S)$, where $f \ S \triangleq \{ f \ x \mid x \in S\}$.

\smallskip

CPOs are closed under product and exponentiation, with orders and bottom elements defined pointwise. We are here interested on exponentiation with sets; i.e., 
if $X$ is a set and $(Y,\preceq,\bot)$ is a CPO then the set $X \to Y$ of functions  from $X$ to $Y$ is a CPO:
 $(X \to Y,\;\lambda\, f f'  \to \forall\, x \in X.\, f\,x \; \preceq \;f'\,x, \; \lambda\,\_ \to \bot)$.  Note the pointwise-defined order and bottom element.
  ACPOs are also closed under exponentation with a set, with same definition as above. 
 The compacts elements of the ACPO $X \to Y$ are functions with compact values and finite support: functions in $X \to Y^\circ$
which evaluate to $\bot$ on all input but finitely many inputs. 

\smallskip

ACPOs result from PPOs  by \emph{completion}. 
For a PPO $(C^\circ,\preceq^\circ,\bot)$ and an ACPO $(C,\preceq,\bot)$, if
 $(C^\circ,\preceq^\circ,\bot)$  is isomorphic to the PPO of compacts of  $(C,\preceq,\bot)$ then  $(C,\preceq,\bot)$ is a \emph{completion of  $(C^\circ,\preceq^\circ)$}; where PPO isomorphisms are bijective, monotonic functions between the underlying sets.

\smallskip

Completions of PPOs exist and are unique up to CPO isomorphisms (where CPO isomorphisms are bijective, continuous functions between the underlying sets; continuity is defined above). For simplicity we mostly use  completions with the property that the underlying PPO isomorphism is the identity.

\smallskip

Moreover, completions also apply to monotonic functions:
if $(D^\circ, \precsim^\circ,\Bot)$ and  $(C^\circ,\preceq^\circ;\bot)$ are PPOs with respective  completions
$(D,\precsim,\Bot)$ and $(C,\preceq,\bot)$, then for any monotonic  $f^\circ : D^\circ \to C^\circ$ there exists a unique continuous  $f : D \to C$ such that $f \ d^\circ = f^\circ \ d^\circ$ for all compacts  $d^\circ \in D^\circ$. We call $f$ the \emph{completion} of~$f^\circ$.
Completions of monotonic functions exist and are unique as well.

\smallskip

  \emph{Kleene's fixpoint theorem} states that
if $F : C \to C$ is continuous then $F$ has the  least fixpoint $\mu F =\lub\,\{F^n\mid n \in \N	\}$, where the sequence $(F^n)_{n \in \N}$ defined by $F^0 = \bot$ and for all $m \in \N$, $F^{m+1} = F (F^m)$.
This can be used for defining functions: by setting $C := A \to B$ with $A$ a set and $B$ a CPO, the least fixpoint of a continuous $F : (A \to B) \to (A \to B)$ defines  a possibly partial function $\mu F : A \to B$.
 We use this technique for 
 encoding  possible partial (co)recursive
  functions in the total language of the Rocq prover. 

\smallskip

 The key condition in Kleene's fixpoint theorem is continuity.
In practice,  this condition is not proved using its definition, but by combining
elementary continuity results~\cite{winskel93}. These results include the continuity of constant and identity functions,  of the composition of continuous functions, and of the  completion of a monotonic function.  
Another result
states that a function from a  product CPO
$X \times Y$ to a CPO $Z$ is continuous in its argument, say, $(x,y)$, if and only if it is continuous in  $x$ and in $y$ separately. 

\smallskip

 %\emph{Knaster-Tarski's fixpoint theorem} gives conditions different from Kleene's fixpoint theorem for the existence of fixpoints. 
 We are also interested in greatest
fixpoints in order to obtain coinductive objects.  
A \emph{complete upper lattice} is a poset $(L,\preceq)$ where $\lub \ S$ exists for all $S \subseteq L$.
 \noindent A consequence of \emph{Knaster-Tarski's  theorem}
says that if $(L, \preceq)$ is a complete upper lattice and $F : L \to L$ is monotonic, then
$\nu F \triangleq \lub\,\{x \in  L  \mid  x \preceq F \ x \}$
is the greatest fixpoint of~$F$.
The specialization of this result to the 
 lattice of subsets of a set $Q$, i.e., $(2^Q, \subseteq)$, entails that any monotonic function $F : 2^Q \to 2^Q$
has a greatest fixpoint $\nu F \subseteq Q$.

\smallskip

We say that 
 $\nu F$  is \emph{coinductively defined} by its \emph{functional} $F$. 
The equation $\nu F = F (\nu F)$ is called the  \emph{unfolding equation of}~$\nu F$.
The fact that $\nu F$ is  the greatest fixpoint of $F$, rewritten as \emph{for all $P \subseteq Q$,   $P \subseteq F\ P$ implies
$P \subseteq \nu F$}, is called the \emph{coinduction principle} of~$\nu F$.

\subsection{Coinductive Types as Labelled Partial Containers}
\label{sec:lpc}
A key ingredient in our domain-theoretical formalization of interaction trees is 
a definition of coinductive types different from the built-in one in Rocq.
In earlier work~\cite{jlamp24} we proposed an
  encoding of coinductive types 
  as completions of inductive types endowed with definition orders. The motivation is that, in Rocq, inductive types are well-supported; by contrast, everything coinductive is quite limited.
  
The resulting  types are kin to (unary) \emph{containers}~\cite{DBLP:journals/tcs/AbbottAG05}, a rich  class of types that include  all \emph{strictly positive} types built with  constants, sums, products, exponentiation by sets, and  fixpoints. 
There are, however, differences: the main one is that our version of coinductive types are ordered, and are inhabited not only by fully defined (maximal) terms, but also by their  partially-defined approximations. 

We recap below the main relevant definitions from  earlier work.
%We use a set-theoretic presentation.
First, the \emph{carrier} of a function $f : X \to Y$, 
where $Y$ contains a value $\bot$, is the pre-image by $f$ of $
Y \setminus\{\bot \}$ and is denoted by $\floor{f}$. 
\begin{definition}[Finite Labelled Partial Container (FLPC)]
  \label{def:flpc}
  Given a set $A$ of \emph{shapes} and two functions $L$ and $B$  that for each $a \in A$ produce a set $(L\ a)$ of \emph{labels} and  respectively 
 a set $(B\ a)$ of \emph{positions},
 the  \emph{finite labelled partial container} (FLPC) $C^\circ$  parameterized by $A$, $L$, and $B$, is the set
  inductively defined by the rules: $\bot \in C^\circ$, and, for all $a \in A$, $l \in L \ a$, and $f : (B \ a) \to C^\circ$
  such that $\floor{f}$ is finite,
  $\node^{\circ} \ a \ l \ f \in C^\circ$. 
\end{definition}    Intuitively, FLPCs are either leaves ($\bot$), or labelled trees of finite depth and of arbitrary-sized breadth, with only finitely many subtrees not being leaves.
(In our earlier work~\cite{jlamp24}
  the trees were unlabelled.)
\begin{definition} [Definition Order]
  \label{def:fcontord}
  The order $\preceq^\circ$ on $C^\circ$ is inductively defined by:
  $\bot \preceq^\circ c^\circ$ for all $c^\circ \in C^\circ$,
  and $\node^\circ \ {a} \ l \ f \preceq^\circ \node^\circ \ {a} \ l \ f'$ whenever
  for all $b \in (B \ a)$, $f \ b \preceq^\circ f' \ b$.
\end{definition}  

\smallskip

\begin{notation}
The above definition order organizes an FLPC parameterized by shapes $A$, labels $L$, and positions $B$,  as a PPO $(C^\circ,\preceq^\circ,\bot)$.
For all $a \in A$, we denote by $F^\circ_{a}$
the set $\{f^\circ : (B \ a )\to C^\circ \mid \floor{f^\circ} \ \mathit{is \ finite}\}$. The relation $\sqsubseteq_a^{\circ} $ on $F^\circ_{a}$
denotes the pointwise order: $f^{\circ}  \sqsubseteq_a^{\circ }f'^\circ $ iff for all $b \in (B\ a) $, $f\ b \preceq^\circ f' \ b$. Let also  $\bot_a$ denote the constant function $\lambda\,\_: (B \ a) \to \bot \in F^\circ_{a}$.
With the above notations, $(F^\circ_a, \sqsubseteq_a^{\circ}, \bot_a)$ is a PPO, and
the functions
$\node^\circ \ {a} \ l : F^\circ_a \to C^\circ$ are monotonic: 
%for all $f^\circ, f'^\circ \in F^\circ_a$,
$f^\circ  \sqsubseteq_a^{\circ} f'^\circ$ implies
$\node^\circ \ {a} \ l \ f^\circ 
\preceq_a \node^\circ \ {a} \ l \ f^\circ$.
\end{notation}

\smallskip

The completion operation (cf.\ Section~\ref{sec:domain})  turns the PPO $(C^\circ,\preceq^\circ,\bot)$ into an ACPO
$(C, \preceq,\bot)$, such that $(C^\circ,\preceq^\circ)$ is isomorphic to
the poset of compacts of $(C,\preceq)$. Hereafter we use a completion in which this isomorphism is the identity.
 We call \emph{Labelled Partial Containers (LPCs)} the completions of FLPCs. 

\smallskip

\begin{notation}
If   $C^\circ$ is an FLPC with shapes $A$, labels $L$, and positions $B$, and the LPC $(C,\preceq,\bot)$ is the  completion of $(C^\circ, \preceq^\circ,\bot)$, then for all $a \in A$,  we  let $F_{a}$  denote the set $(B \ a) \to C$. A relation   $\sqsubseteq_a$ on $F_a$
is defined by $f  \sqsubseteq_a f' $ iff for all $b \in (B\ a) $, $f \ b \preceq f' \ b$. 
\end{notation}

\medskip

Then, by the closure of ACPOs under exponentiation with a set, $(F_a, \sqsubseteq_a,\bot_a)$ is an ACPO whose PPO of compacts is $(F^\circ_a, \sqsubseteq^\circ_{a}, \bot_a)$, which amounts to saying that $(F_a, \sqsubseteq_a,\bot_a)$  is a completion of $(F^\circ_a, \sqsubseteq^\circ_{a}, \bot_a)$. Hence, from each monotonic 
function $\node^\circ \ {a} \ l : F^\circ_{a} \to C^\circ$ one obtains,
by function-completion, a unique continuous function $\node \ {a} \ l : F_a \to C$
satisfying $\node \ a \ l \ f^\circ_{a} = \node^\circ \ a \ l \ f^\circ_{a}$ for all
$f^\circ_{a} \in  F^\circ_a$. Moreover, the functions $\node^\circ \ a \ l$ behave as constructors
for $C$, which, in turn, behaves like  a coinductive type\footnote{Countably infinite terms, illustrating the coinductive nature of $C$, can be built using Kleene's fixpoint theorem.
}:

\begin{itemize}
\item  for all $c \in C$, $c =\bot$ or (exclusively) there exist unique $a\in A$, $l \in L \ a$ and $f \in F_a$ such that $c = \node \ a \ l \ f$; 

\item using Knaster Tarski's fixpoint theorem and associated notions we coinductively define a relation % \emph{bisimulation relation} 
$\precsim \;\subseteq C \times C$, whose unfolding equation states that $ c \precsim c' $ iff
 $c = \bot$ or there exist $a\in A$,  $l \in L \ a$  and $f,f'\in F_a$ such that $c = \node \ a \ l \ f$, $c' = \node \ a \ l \ f'$ and, for all $b \in (B\ a), (f \ b)  \precsim  (f'  \ b)$. Moreover, we prove that 
 $\precsim$ coincides with the definition order $\preceq$;

\item the order $\precsim$ gives rise to a \emph{bisimulation relation} $\approx\;\subseteq C \times C$, defined by $t \approx t'$  iff $t\precsim t'$ and $t' \precsim t$, whose \emph{unfolding equation} states that $ c \approx c' $ iff
 $c = c' = \bot$ or there exist $a\in A$,  $l \in L \ a$  and $f,f'\in F_a$ such that $c = \node \ a \ l \ f$, $c' = \node \ a \ l \ f'$ and, for all $b \in (B\ a), (f \ b)  \approx (f'  \ b)$;

 \item  bisimulation coincides with equality:  to prove $c = c'$ one can equivalently prove $c \approx  c'$;
 
 \item the bisimulation relation satisfies the following \emph{coinduction principle}: in order to prove $c \approx  c'$, find $R \subseteq C\times C$ with $(c,c') \in R$ and prove  that for all $(u,v) \in R$, $u = v =  \bot$ or $u= \node \ a \ l \ f$, $v = \node \ a \ l \ f'$ for some $a \in A$, $l \in L \ a$ and $f,f' \in F_a$ such that for all $b \in (B \ a)$, $((f \ b),(f'  \ b)) \in R$.

\end{itemize}
All these results are inherited by interaction trees, defined later in the paper as LPCs with specific shapes, labels and positions.

\section{CPO-valued Monads}
%\vlad{Je pense que les CPO-monades méritent leur propre section}
\label{sec:monads}
%\vlad{je garderais juste CPO-monad}
%\subsubsection{CPO-monad}
A \textit{CPO-valued monad} is a monadic structure whose values are organized as a Complete Partial Orders. 
All the monadic structures occurring in this paper, including interaction trees, are CPO-valued monads. 
%We note that similar notions, with variations (strictness of $\bind$, $\omega$-CPOs instead of CPOs,\ldots) have recently been introduced~\cite{DBLP:conf/ecoop/DagninoGZ25,DBLP:conf/esop/GavazzoTV24,DBLP:journals/pacmpl/Kavvos25}.

\begin{definition}
\label{def:cpo_monad}
A CPO-valued monad is a structure $(M, \ret, \bind)$ where
$M: \Set \to \CPO$ is a functor from sets to CPOs, and,
for each set $X$ and $Y$,
$\ret_X: X \to M \ X$ and
$\bind_{X,Y}: M \ X \to (X \to M \ Y) \to M \ Y$ are functions such that,
with the  notation
$
\mdo{x \leftarrow m; \; m'} \triangleq \bind_{X,Y}\, m\, (\lambda x.\, m')
$:
\begin{itemize}

\item $\forall X \; Y \; (k: X \to M \ Y),\ \lambda m.\,\mdo{x \leftarrow m; k\, x}$ is a continuous function between CPOs $M \ X$ and $ M \ Y$;

\item $\forall X \; Y \; (m: M \ X),\ \lambda k.\,\mdo{x \leftarrow m; k\, x}$ is a continuous function between CPOs $(X \to M \ Y)$ \footnote{here we use the fact that CPOs are closed under exponentiation with a set.} and $ M \ Y$.

\item $\ret$ and $\bind$ satisfy the standard monadic laws:
\begin{itemize}

\item $\forall X \; (m: M\ X),\; \mdo{x \leftarrow m; \; \ret\, x} = m$

\item $\forall X \; Y \; (x: X) (f: X \to M\ Y),\ \mdo{x' \leftarrow \ret_X \, x; \; f \, x'} = f\, x$

\item
$
\forall X \; Y \; Z \; (m: M\ X) (f: X \to M\ Y) (g: Y \to M\ Z),\\
\mdo{y \leftarrow (\mdo{x \leftarrow m; \; f\, x}); \; g\, y} = \mdo{x \leftarrow m; \; \mdo{y \leftarrow f\, x; \; g\, y}}$.
\end{itemize}
\end{itemize}
\end{definition}

\begin{remark}
CPO-valued monads diverge from the usual definition of monads: their functors are not endofunctors, but functors from sets to CPOs\footnote{the effect of the functor on morphisms: by the continuity 
of $\bind$ in its first argument, every function $f:X  \to Y$ is mapped by the functor $M$ to $M \ f :=
\lambda\,m \Rightarrow \mdo \ x \leftarrow m\;; \ret \ (f \ x)$, which is a  continuous function from $M \ X$ to $M \ Y$.}. Attempts to cast our definition in the proper categorical definition of monads, i.e., requiring endofunctors between CPOs, results in more implementation work (complications due to dependent types, more continuity proofs, \ldots) without much practical benefit. Hence we pragmatically choose the  hybrid notion of CPO-valued monads. 
\end{remark} 
\begin{notation}
   For sets $X$ and $Y$, $X+Y$ denotes their disjoint sum, with injections 
    $\inl : X \to X+Y$ and $\inr : Y \to X + Y$. 
   Hereafter, $\mathbf{1}$ denotes the unit 
    set, with exactly one element, $\ttt$.
\end{notation}

\begin{example}[The option CPO-valued monad] This monad  consists of:
\label{ex:option}
\begin{itemize}
\item the functor $\mathit{option}$, which  maps
any set $X$ to the flat CPO
$X \cup \{\bot\}$;
\item for all sets $X$ and $x \in X$, $\ret \ x = x$;
\item $\mdo x \leftarrow m; \, f \, x$ is $\bot$ if $m =\bot$ and $f \, m$ otherwise.
\end{itemize}
The continuity requirements of a CPO-monad
are proved using the techniques mentioned in Section~\ref{sec:domain}.
\end{example}
Next, we define the state CPO-valued monad transformer.
A CPO-valued monad transformer is defined to augment an existing CPO-valued monad with a new effect, producing another CPO-valued monad. This structure also provides a method to lift operations from the original monad into the augmented one. 
We shall be using the state monad transformer in combination with the option monad from Example~\ref{ex:option}.

\begin{example}[The state CPO-valued monad transformer]
\label{ex:statet}
Given a CPO-valued monad $(M, \ret_M,$ \linebreak $\bind_M)$, the state CPO-valued monad transformer parameterized by a set $S$ of states constructs a new CPO-valued monad $(\statet_S^M, \ret, \bind)$ as follows:
\begin{itemize}

\item
$\statet_S^M \; X$ maps each set $X$ to the CPO $M(X\times S)$.

\item
$\ret \; x$ is defined as $\lambda s . \; \ret_M (x, s)$. It produces a stateful computation that leaves the state unchanged and yields the value $x$ inside the underlying monad $M$.

\item
$\mdo{x \leftarrow m; \; f \; x}$ is defined as $\lambda s . \; \mdo{(x, s') \leftarrow m \; s; \; f \; x \; s'}
$. This operation sequences two stateful computations, threading the state through both while remaining inside the context of $M$.

\end{itemize}
This CPO-valued monad transformer is equipped with two fundamental primitives, $\get$ and $\sput$:
\begin{itemize}
\item
 $\get$ is defined as $\lambda s . \; \ret_M (s, s)$.
It takes a state $s$ and returns a computation in $M$ that produces the pair $(s, s)$,  yielding the current state value while leaving the state  unchanged.

\item $\sput \; s'$  is defined as $\lambda s . \; \ret_M (\ttt, s')$.
This operation ignores the incoming state $s$ and returns a computation that produces the pair $(\ttt, s')$. This signifies that the operation's result is the unit value $\ttt$ (a placeholder indicating no meaningful result), and the state is updated to $s'$.

\end{itemize}
We discharge the continuity requirements monads by applying the techniques mentioned in Section~\ref{sec:domain}.
\end{example}
By instantiating the state CPO-valued monad of state-set $S$ with the option monad we obtain 
a CPO-valued  monad combining state and option effects; its functor $\mathit{State}_S$ maps each set $X$ to the flat CPO $(X \times S) \cup\{\bot\}$. This is used later in the paper, in the Syracuse example.
\section{Interaction Trees}
\label{sec:itree}
We now focus on interaction trees: their definition and basic properties, and 
their monadic structure.
\subsection{Interaction Trees: Definition and Basic Properties}
\label{sec:itreecont}

We  define interaction trees as instances of Labelled Partial Containers (cf. Section~\ref{sec:lpc}). This provides interaction trees with a coinductive structure and order relation, and a bisimulation relation equivalent to equality.
The encoding is quite involved but in practice (i.e., in Rocq) it is hidden: users  manipulate interaction trees without having to know how they are encoded. We start with \emph{finite} interaction trees and their encoding as FLPCs, and then we obtain the general  interaction trees using completion.

\begin{definition}[Finite Interaction Tree]
A \emph{finite interaction tree} $\itree^\circ \ F \ X$ parameterized by a partial function $F : \Set \to \Set$ (the \emph{effect signature}) and a set $X$ is inductively defined by the rules 
\begin{itemize}
\item $\bot \in \, \itree^\circ \ F \ X$;
\item for all $x \in X$, $\pure_{X}^\circ \ x \in \,\itree^\circ \  F \ X$;
\item for all $Y \in \Set$, $\mathit{fy} \in F\ Y$, and $k : Y \to \itree^\circ \ F \ X$
such that $\floor{k}$ is finite,
 $\impure_{Y}^\circ \ \mathit{fy} \ k \in \, \itree^\circ \ F \ X$.
\end{itemize}
\end{definition}

\begin{remark}[Size Issue]
Although we speak informally of $\Set$ and ``partial functions" from $\Set$ to $\Set$, we implicitly work in a suitable Grothendieck universe (or a hierarchy of universes) to avoid size issues. This is handled formally in our Rocq development via Rocq's template polymorphism (universe polymorphism), which manages universe levels precisely. The mathematical presentation suppresses these universe annotations to highlight the conceptual structure rather than the foundational bookkeeping.
\end{remark}

Next, we identify the instance of FLPC (cf.~Def.~\ref{def:flpc}) that 
encodes finite interaction trees:

\begin{definition}
For  $F : \Set \to \Set$ and $X \in \Set$,  we define an FLPC
${\cal L}^\circ \ F \ X$
with the following parameters:
\begin{itemize}
\item the of shapes set $A$  is the disjoint union
$\mathbf{1} + \Set$;
\item  $B: A \to \Set$  is defined by 
$B \ (\inl \ \ttt) = \emptyset$ and $B \ (\inr \ Y) = Y$ for all $Y \in \Set$;
\item $L : A \to \Set$  is defined by
$L \   (\inl \ \ttt)  = X$ and $L \ (\inr \ Y)  = F \ Y$  for all $Y \in \Set$.
\end{itemize}
The constructors of $\itree^\circ \ F \ X $ are encoded as using the constructors of 
${\cal L}^\circ \ F \ X$ as follows:
\begin{itemize}
\item $\bot \in \itree^\circ \ F \ X $ is encoded by $\bot \in {\cal L}^\circ \ F \ X$;
\item for all $x \in X$, $\pure_{X}^\circ \ x$  is encoded by $\node^\circ  \ (\inl \ \ttt) \ x \ (\lambda\,\_:\emptyset  \to \ X)$, where $\lambda\,\_:\emptyset  \to \ X$ is the unique element of\,
$\emptyset \to X$;
\item for all $Y \in \Set$, $fy \in F \ Y$ and $k \in Y \to \itree^\circ \ F \ X$ with 
$\floor{k}$ finite, $\impure_{Y}^\circ \ \fy \ k$ is encoded by $\node^\circ  \ (\inr \ Y) \ \fy  \ k$. 
\end{itemize}
\end{definition}
By applying completion to ${\cal L}^\circ \ F \ X$ with its definition order, we obtain the LPC ${\cal L}\ F \ X$. The  encodings of the functions $\impure_{Y}^\circ \ \fy \ k$ are monotonic in $k$ and are completed to functions  $\impure_{Y} \ \fy \ k$ continuous in $k$. The functions $\pure_{X}^\circ$ are unchanged but, for notation consistency, are renamed  $\pure_{X}$.

\begin{definition}[Interaction Trees]
\label{def:itree}
Given
a partial  function $F : \Set \to \Set$  and a set $X$, $\itree \ F \ X$ is
 defined as ${\cal L}\ F \ X$.
\end{definition}

\begin{remark}
Thanks to the general properties of LPCs (cf. Section~\ref{sec:lpc}), for all $t \in \itree \ F \ X$, $t =\bot$ or (exclusively) there exists a unique 
 $x \in X$ such that $t = \pure_{X} \ x$, or (exclusively) there exist unique 
 $Y \in \Set$, $\fy \in F \ Y$, and $k : Y \to \itree \ F \ X$ such that $t = \impure_Y\ \fy \ k$.
This can be understood as follows: $\itree \ F \ X$ is a set of labelled trees of arbitrary breadth and possibly infinite depth, whose leaves are either $\bot$ or $\pure_{X} \ x$ for some $x \in X$,
and nodes are of the form $\impure_Y \ \fy \ k $ for $Y\in \Set$, where $\fy\in F \ Y$ is the node's label and $k: Y \to \itree \ F \ X$ defines  the set of subtrees (of cardinality $\mid Y \mid$) of the node. Infinite depth is obtained by nesting the $\impure$ constructor a countably infinite number of times. 

The encoding of interaction trees as LPCs ensures that $(\itree \ F \ X,\preceq,\bot)$ is an ACPO, where $\preceq$ is the completion of the definition order $\preceq^\circ$ of ${\cal L}^\circ \ F \ X$.
\end{remark}

 We now introduce a coinductively defined relation $\precsim$:

%The completion $\preceq$ of the definition order $\preceq^\circ$ over
%$\itree^\circ \ F \ X$ is proved to be equivalent to a coinductively defined 
%relation:
\begin{definition}
\label{def:coindr}
For all $X \in \Set$, the  relation $\precsim\;\subseteq  \itree \ F \ X \times  \itree \ F \ X$ 
is coinductively defined as follows:
$t \precsim t'$ iff either
 $t = \bot$, or $t = t' = \pure_{X} \ x$ for  some $x \in X$, or there exist $Y \in \Set$,  $\fy \in F \ Y$, and $k,k' :  Y \to \itree \ F \ X$ such that $t  = \impure_Y  \ \fy \ k \ $, $t' = \impure_Y \ \fy \ k'$ and for all $y \in Y, (k \ y)  \precsim (k'  \ y)$.
 \end{definition}

General properties of LPCs (cf. Section~\ref{sec:lpc}) imply that  $\precsim$ from Definition~\ref{def:coindr} is equal to  
 $\preceq$: in particular, $\precsim$ is an order.
%equivalent and $(\itree \ F \ X,\preceq,\bot)$ is an ACPO
%(Algebraic CPO).
 %\begin{remark}
%The coinductive definition of order is based on a corollary to the Knaster-Tarski  theorem
%(cf.\  section~\ref{sec:tarski}). Again, this is not the builtin coinduction of Rocq.
% \end{remark}
The $\precsim$  order then induces an equivalence relation:

\begin{definition}[Bisimulation]
\label{def:coind}
The bisimulation relation $\approx\;\subseteq  \itree \ F \ X \times  \itree \ F \ X$ 
is defined by $t \approx t'$ iff $t \precsim t' \wedge t' \precsim t$.
 \end{definition}
 Since $\precsim$ is an order, we  obtain by reflexivity and antisymmetry:

 \begin{lemma}
 \label{lem:eqcoind}
 Equality on interaction trees coincides with bisimulation:  $t = t'$ iff $t \approx  t'$.
 \end{lemma}
Thanks to the general properties of LPCs (cf. \ Section~\ref{sec:lpc}) we also obtain the following result, which, in combination with Lemma  \ref{lem:eqcoind},
 is used for proving equalities of interaction trees:

 \begin{lemma}[Coinduction Principle for Bisimulation]
 \label{lem:coind}
 Given $t,t' \in \itree \ F \ X$, in order to prove $t \approx  t'$  it is enough  
 to find $R \subseteq \itree \ F \ X \times  \itree \ F \ X $ with $(t,t') \in R$ and prove  that for all $(u,v) \in R$, $u = v =  \bot$ or 
 $u = v = \pure \ x$ for some $x \in X$, or $u  = \impure_Y  \ \fy \ k $, $t' = \impure_Y \ \fy \ k'$ for some  $Y \in \Set$, $\fy \in F \ Y$, and $k,k' :  Y \to \itree \ F \ X$, 
 such  that  for all $y \in Y, ((k \ y), (k'  \ y)) \in R$.
\end{lemma}

We now give %two examples 
an example of effect signatures for interaction trees that are used ahead in the paper. Remember that  interaction trees  do not interpret their effects.

% \begin{example}[Probability effect signature]
% \label{ex:prob_effect}
% The signature of probabilistic effects is defined by a partial function  $\mathit{Prob} : \Set \to \Set$  defined only on Booleans:
% $\mathit{Prob} \; \mathbb{B} \; = \; \bigcup_{p \in [0,1] } \; \{\flipeff \; p\}$.
% %
% This defines an interface for probabilistic computations: the only  primitive effect is $\flipeff \; p$ and represents a probabilistic choice.
% %
% The interaction trees with effect signature  $\mathit{Prob}$  can thus represent programs that make repeated probabilistic choices, branching according to the outcomes of these choices.
% \end{example}

\begin{example}[State effect signature]
\label{ex:state_effect}
The signature of state-manipulating effects over a set $S$ of states  is defined by a partial function $\mathit{State}_S : \Set \to \Set$, defined only on the sets $S$ and $\mathbf{1} $ by:
$$
\begin{array}{l}
\mathit{State_S} \; S \;\; = \;\; \{ \geteff \} \\
\mathit{State_S} \; \mathbf{1} \;\; = \;\; \bigcup_{s \in S } \; \{\sputeff \; s \} %\\

\end{array}
$$
This defines the interface for stateful computations. %The primitive effects are: $\geteff$, for reading from the current state; and $\sputeff \; s$ for setting the state to $s$.
The corresponding interaction tree can thus represent programs that interact with a mutable state cell by reading from and writing to it over the course of their execution.
\end{example}

The following function takes an effect $\fy$ of type $F \; Y$ turns it into an elementary interaction tree:
\begin{definition}[Trigger]
\label{def:trigger}
$$
\trigger \; \fy \;\; = \;\; \impure_Y \; \fy \; \pure_Y.
$$
\end{definition}
Intuitively, $\trigger$ is the most basic way to inject a single, primitive effect into an interaction tree for the corresponding effect.
It takes an uninterpreted effect $\fy$ (like %$\flipeff \; p$ or
$\geteff$) and wraps it into an interaction tree. The resulting tree is a single effect node whose immediate continuation is $\pure$.

 Interaction trees have one additional primitive: $\fold$, which is mainly used  for interpreting the effects contained in a tree
 and accumulating them into a single value. It is presented in the next section because its definition shares some features with the $\bind$ connector, part of the CPO-valued monad structure.
 
 \subsection{Interaction Trees as CPO-valued Monads}
 We now show that interaction trees are instances of the CPO-valued monads from Definition~\ref{def:cpo_monad}.
 We start by defining $\ret$ and $\bind$  and prove the continuity of $\bind$. Specifically, $\bind$ is a corecursive function, defined using Kleene's theorem as the least fixpoint of its continuous functional. The monad laws are proved using the coinduction principle for interaction trees (cf.~Lemma~\ref{lem:coind}). Thanks to the equivalence between equality and bisimulation from Lemma~\ref{lem:eqcoind}, the monad laws are Rocq equalities\footnote{This in contrast to \cite{itree19}, where monad laws only hold up to a certain weak bisimulation.}. 
 
 Then, we define a $\fold$ function for interaction trees, also as the least fixpoint of its continuous functional, and prove its continuity. $\fold$ is used, among others, for traversing interaction trees while  interpreting and accumulating the effects therein, which endows interaction trees with \textit{denotations}.

 The section continues with our definition of program equivalence and its correctness property.
 
 \subsubsection{Ret, Bind, and Monadic Laws}
 \label{sec:fbind}
 For interaction trees parameterized by an effect signature $F$
 and a set $X$, we define the monadic elementary computation $\ret_X : X \to \itree \ F \ X$  by $\ret_X = \pure_X$. For the sequencing of computations
 $\bind : \itree \ F \ X \to (X \to \itree \ F \ Y) \to \itree \ F \ Y$, we first identify its expected properties:
 \begin{itemize}

 \item for all $f : X \to \itree \ F \ Y$, $\bind \ \bot \ f = \bot$ ;
 \item for all $x \in X$ and $f : X \to \itree \ F \ Y$, 
 $\bind \ (\ret_X\ x) \ f =  f \ x$;
 \item for all $Z  \in \Set$, $\fz \in F \ Z$, $k : Z \to \itree \ F \ X$,
 and  $f : X \to \itree \ F \ Y$,
 $\bind \ (\impure_Z \ \fz \ k) \ f $ = $\impure_Z  \ \fz  \ (\lambda\,z. \ \bind \ (k \ z) \ f)$.
 \end{itemize}
 The first property states that sequencing an undefined computation with anything produces an undefined computation; the second property is one of the monad laws (left neutrality of $\ret$ w.r.t.\ $\bind$); and the third property defines
 $\bind$ in terms of itself.
 % If interaction trees were defined using the builtin coinduction of Rocq, one could define $\bind$ using a Rocq corecursive function %(a \textit{CoFixpoint})
 % that amounts to stating the three above  identities. But interaction trees are not defined using the coinduction of Rocq. Hence, 
 In order to define $\bind$ we use Kleene's fixpoint theorem.
 Since Kleene's theorem can only define functions of one argument, but $\bind$ has two arguments, we shall first define its \emph{uncurried} version $\widehat{bind} : 
 (\itree \ F \ X) \times (X \to \itree \ F \ Y) \to \itree \ F \ Y$. The corresponding functional $\widehat{\fbind}$ for effect signature $F$ and sets $X,Y$ is written using pseudocode in Figure~\ref{funcbind}. Its type is
 $((\itree \ F \ X) \times (X \to \itree \ F \ Y ) \to \itree \ F \ Y ) \to
 ((\itree \ F \ X) \times (X \to \itree \ F \ Y ) \to \itree \ F \ Y )$. 
 
 We note that the domain and codomain of $\widehat{\fbind}$ are the same and are CPOs, because interaction trees are CPOs and 
 CPOs are closed under exponentiation and product. We prove that $\widehat{\fbind}$ is continuous by the  techniques mentioned in Section~\ref{sec:domain}. Then, Kleene's theorem ensures that
 $\widehat{\fbind}$ has a least fixpoint, of type $(\itree \ F \ X) \times (X \to \itree \ F \ Y ) \to \itree \ F \ Y$. We call  this least fixpoint $\widehat{\bind}$.
 By \emph{currying} $\widehat{\bind}$ we obtain the desired  $\bind$ function, of the proper type. By expanding the fixpoint equation $\widehat{\fbind} \ \widehat{\bind} = \widehat{\bind}$ we obtain the identities from the beginning of this section.

 \begin{figure}[ht]
\noindent$\widehat{\fbind}$
    $(\phi : (\itree \ F \ X) \times (X \to \itree \ F \ Y ) \to \itree \ F \ Y )$
     $(p: \itree \  F  \ X \times (X \to \itree \ F  \ Y ))$ : 
     $\itree  \ F  \ Y$:=

    $\llet \ (m,f) = p\ \iin$
    
    \quad $\ccase \ m \ \oof$
    
     \quad \quad $\bot \Rightarrow \bot$

     \quad \quad $\pure_X \ x \Rightarrow f \ x$     

     \quad \quad $\impure_Z \ \fz \ k \Rightarrow \impure_Z \ \fz \ (\lambda\,z. \ \phi \ ((k \ z),f))  $
     
\quad $\eend$
\caption{\label{funcbind} Functional $\widehat{\fbind}$  for the uncurried $\widehat{\bind}$.}     
\end{figure}

In order to conform to Definition~\ref{def:cpo_monad} of CPO-valued monads we still need to prove that $\bind$ is continuous in both arguments and that the monad laws hold.

\begin{itemize}
    \item Proving that $\bind$ is continuous in both arguments is equivalent, by a result mentioned in Section~\ref{sec:domain}, to proving that $\widehat{\bind}$ is continuous in its single argument (a pair). We proceed as follows:
    \begin{itemize}
        \item We first prove that  $\widehat{\fbind}$ preserves continuity: if $\phi:  (\itree \ F \ X) \times (X \to \itree \ F \ Y ) \to  \itree \ F \ Y $ is continuous, then $\widehat{\fbind} \ \phi: (\itree \ F \ X) \times (X \to \itree \ F \ Y ) \to  \itree \ F \ Y$ is continuous as well;
        \item Then, we prove by induction over natural numbers that for all $n\in \mathbb{N}$, the $n$-th iteration of  
        $\widehat{\fbind}$ starting with a constant $\bot$ function, i.e., 
        $\widehat{\fbind}^{(n)} (\lambda.\_ \Rightarrow \bot)$, is continuous;
        
        \item Finally, we note that the set $\{ \widehat{\fbind}^{(n)} (\lambda.\_ \Rightarrow \bot)\mid n\in \mathbb{N}\}$ is directed, and  use a result from domain theory stating that the $\lub$ (here, $\widehat{\bind}$) of a directed set of continuous functions  is a continuous function as well. This ensure that $\widehat{\bind}$, and then ${\bind}$, are continuous.
    \end{itemize}
    \item For the monad laws: the fact that $\ret$ is neutral to the left of $\bind$
    is already the second identity given at the beginning of the section. The two other laws - $\ret$ is neutral to the right of $\bind$, and $\bind$ is associative - are stated as equalities, transformed thanks to Lemma~\ref{lem:eqcoind} into bisimulations, and proved using the coinduction principle for bisimulation on interaction trees (Lemma~\ref{lem:coind}).
\end{itemize}

 \subsubsection{Fold}
 \label{sec:ffold}
 %\vlad{trigger is also a primitive}
Interaction trees have one additional primitive: $\fold$, for traversing a tree and accumulating some function of its nodes into one value. This is used for
computing the accumulated effects encoded in a tree, which amounts to endowing interaction trees with a denotational semantics.

Since interaction trees can be  partially defined (i.e., some leaves are $\bot$), their denotations can be partially defined as well. To account for such situations we let the return value of $\fold$ inhabit a CPO.

We first give the expected properties of the $\fold$ function, then define it using Kleene's fixpoint theorem.
Given $F : \Set \to \CPO$ (the effect signature), a set $X$, a CPO $Y$,
a function $f : X \to Y$, and a function $g_Z : F \ Z \to (Z \to Y) \to Y$
parameterized by a set $Z$, $\fold \ f \ g_Z : \itree \ F \ X \to Y$ is such that
\begin{itemize}
\item $\fold \ f \ g_Z \ \bot = \bot$;
\item $\fold \ f \ g_Z \ (\pure_X \ x) = f \ x$;
\item $\fold \ f \ g_Z \ (impure_Z \ \fz \ k)   = 
g_Z \ \fz \ (\lambda\,z. \ \fold \ (k \ z) )$.
\end{itemize}

Consequently, the expected functional for $\fold$ is given below.

\begin{figure}[ht]
\noindent$\ffold$
    $(f : X \to Y )$
     $(g_Z : F \ Z \to (Z \to Y) \to Y)$ 
     $(\varphi :  \itree  \ F \ X \to Y)$
     $(t : \itree  \ F \ X)$ : $Y$ :=

    \quad $\ccase \ t \ \oof$
    
     \quad \quad $\bot \Rightarrow \bot$

     \quad \quad $\pure_X \ x \Rightarrow f \ x$     

     \quad \quad $\impure_Z \ \fz \ k \Rightarrow g_Z \ \fz \ (\lambda\,z. \ \varphi \ (k \ z))$
     
\quad $\eend$
\caption{\label{fig:funcfold} Functional for $\fold$.}     
\end{figure}

However, there is a catch. Proving that
$\ffold \ f \ g_Z : (\itree  \ F \ X \to Y) \to (\itree  \ F \ X \to Y) $
is continuous (as required by Kleene's fixpoint theorem) requires the  hypothesis that for all 
$Z \in \Set$ and $\fz : F \ Z$, $(g_Z \ \fz) : (Z \to Y) \to Y$ is continuous as well. In the dependently-type setting of Rocq this amounts to adding an argument to $\ffold$: a proof of continuity for $g_Z \ \fz$. Here we omit the additional argument for the sake of readability. Using this additional hypothesis, the proof of continuity of $\ffold \ f \ g_Z$ is performed using the techniques mentioned in Section~\ref{sec:domain}. By Kleene's fixpoint theorem, the least fixpoint of $(\ffold \ f \ g_Z)$ is 
$\fold \ f \ g_Z$, which satisfies the requirements stated above for the $\fold$ function. Moreover, we prove that $\fold \ f \ g_Z$ is continuous as a function from the CPO $\itree  \ F \ X \to Y$ to itself, by employing the same techniques  from Section~\ref{sec:fbind} that were used for the continuity of $\widehat{\bind}$.

We now give an example of using $\fold$ for giving denotations to interaction trees.
This is achieved  by calls of the form $\fold \; \ret \; g \; m$, which traverses the interaction tree $m$ and applies the function $g$ to each effect in $m$.  The overall result is a value in a concrete monad, typically, one implementing the effect of~$m$.

\begin{example}[Denotation of the interaction-tree state monad]
Given an underlying monad $M$,
the denotations induced by the parameter $g$ of $\fold$, i.e., $g \; \geteff$ and  $g \; (\sputeff \; s')$ into the state monad transformer $\mathit{StateT}_S^M$ (cf. Example~\ref{ex:statet}) are given respectively by $\lambda k .  k \ s \ s$ and $\lambda k .  k\ttt   s'$.
Intuitively, these definitions describe how to handle state operations:
\begin{itemize}
    \item For $\geteff$: The function $g$ produces an action that takes the current state $s$. It then calls the function $k$ with two arguments: first, the value $s$ (so the rest of the computation can use it), and second, the state $s$ again (leaving the state unchanged for the subsequent computation).
    \item For $\sputeff \; s'$: The function $g$ produces an action that takes a new state $s'$. It then calls the function $k$, passing a dummy unit value $\ttt$ (since a put operation carries no meaningful result) and, crucially, updating the state to $s'$ for all future operations.
\end{itemize}
In this way, $g$ translates the abstract $\geteff$ and $\sputeff \; s'$ effects into concrete state-passing actions.
\end{example}

\subsection{Congruence for Equational Reasoning on Interaction Trees}
\label{sec:congruence}
 
We now turn to defining congruence relations for interaction trees. First, why does one
need congruences in the first place? Why is bisimulation (which, in the present case, coincides with equality) not good enough? The reason is the very nature of interaction trees, as structures that sequence  and combine effects but do not interpret them. The interpretation is the task of the \emph{denotation} operation, which gives us an important criterion for a congruence to be 
\emph{correct}: if two interaction trees are congruent, then that the two trees must have the same denotation.

Although we do not prove the converse (completeness), the relation is sufficiently powerful for proving the congruence of non-trivial programs (as will be illustrated in Section~\ref{sec:examples}).

Our  relation $\equiv$ is actually the union of a family of relations
parameterized by an effect signature $F : \Set \to \Set$. Each relation $\equiv_F$ in the family
is such that $\equiv_F \subseteq \bigcup_{X\in \Set} (\itree F \ X) \times (\itree F \ X)$; that is,
  $\equiv_F$
relates trees in $\itree F \ X$, for all sets $X$. Each $\equiv_F$ is defined as the \emph{least}
relation 
satisfying the following constraints:
\begin{itemize}
\item first, $\equiv_F$ includes  a \emph{base} relation, specific to the effect signature $F$. Intuitively, the base relation encodes axioms satisfied by combinations of primitive effects. It is the only part of the $\equiv_F$ relation  that is specific to $F$: we sometimes say that the base relation \emph{generates} $\equiv_F$:
\item then, $\equiv_F$ is closed under reflexivity, symmetry, and transitivity;
\item next, in order to take into account the CPO  structure of interaction trees, $\equiv_F$ is a congruence w.r.t.\ $\lub$. Specifically, for any two directed sets $S$, $S'$ such that \emph{for each $s \in S$ there exists $s' \in S'$
with $s' \equiv_F s$}, and conversely, \emph{for each $s' \in S'$ there exists $s \in S$
with $s \equiv_F s'$}, we have $\lub\ S \equiv_F \lub\ S'$;
\item finally, in order to take into account the CPO-valued monad structure of interaction trees, $\equiv_F$ is a congruence w.r.t.\ $\bind$. Specifically, for all  $X,Y \in \Set$ $m,m' \in \itree \ F \ X$ and $f,f' : X \to \itree \ F \ Y$
if $m \equiv_F m'$ and for all $x \in X$, $f \ x \ \equiv_F f' \ x$ then $\bind \ m \ f \equiv_F \bind \ m' \ f'$.
\end{itemize}
\begin{remark}
We emphasize that each $\equiv_F$, being the smallest relation satisfying the above constraints, is an \emph{inductively} defined relation. An inductively-defined relation has the advantage of enabling proofs by induction; for example, the proof of correctness of $\equiv$ critically depends on  induction being available\footnote{Specifically, we have designed a customized induction principle for $\equiv$ and have proved it sound in Rocq.}.
A potential drawback of an inductive relation for interaction trees modeling programs is that each pair of programs must be proven equivalent in \emph{finitely} many steps; this may appear to rule out equivalence proofs for \emph{infinitely-running}, i.e., non-terminating programs, be they recursive or corecursive. This pitfall is avoided by $\equiv_F$ being a congruence for $\lub$: non-terminating programs are the $\lub$s of their finite approximations; the congruence to $\lub$ thus reduces the equivalence of non-terminating programs to equivalences of their respective finite approximations, which are adequately handled by induction.
\end{remark}

\begin{example}
\label{ex:basestate}
The congruence relation for the effect signature $F = \mathit{State}_S$
(Example~\ref{ex:state_effect})
is generated by the base relation
$\{((\sputeff \ x;\; \sputeff \ y) , \sputeff \ y), \ ((\sputeff\ x;\; \geteff),(\sputeff \ x;\; \ret \ x)), \ ((\mdo \ x \leftarrow \geteff ;\; \sputeff \ x), (\ret \ *)), \ (\mdo x \leftarrow \geteff;\; \mdo \ y \leftarrow \geteff;\; f\  x \ y), (\mdo \ x \leftarrow \geteff;\; f\  x \ x))\}$, implicitly universally quantified over the free variables.
Here, $\ret$ and $\bind$ (under the \textit{do} notation) are those of the interaction-tree state CPO monad $\itree \ \mathit{State}_S \ X$, with  $*$ being the unique value in the unit set $\textbf{1}$.
\end{example}

\section{Example}
\label{sec:examples}

The main advantage of our interaction-tree monad is that it allows one to reason equationally about the program's syntax independently of the effectful semantics given later by an interpreter. This separation is absent in standard monads, where the program's logic and its execution are fused.

We illustrate this on 
%two examples. The first one is 
the equivalence of two possibly non-terminating programs using the $\mathit{State_S}$ effect signature. 
%The second one
%establishes the correctness of a possibly non-terminating program using the combined  $\mathit{State_S}$  and $\mathit{Prob}$ effect signature, by proving it equivalent to a more abstract reference implementation.
%\subsection{Syracuse}
The Syracuse problem is a well-known open mathematical conjecture about the finiteness of a sequence of natural numbers. We do not, of course, solve the conjecture, but merely show that two programs enumerating the elements in the sequence in two different ways are equivalent for the congruence $\equiv_{\mathit{State_S}}$. By the correctness of 
this  relation, the denotations of the two programs are equal; in particular, the denotations either both terminate with the same output, or they both diverge.

\begin{figure}[ht]
\centering
\begin{minipage}{.5\textwidth}
\noindent$\mathit{syracuse\_single}(n:\mathbb{N}) :=$

\quad $\sputeff \ n;$

\quad $ \while \ (\mdo \ x \leftarrow 
\geteff;\; \ret \ (x \not = 1) )$ 

\quad \quad $\{\{$

\quad \quad \quad $\mdo \ x \leftarrow \geteff;$

\quad \quad \quad $\iif \  \even \ x  \ \tthen$

\quad \quad \quad \quad $\sputeff \ (x \ \mathit{div} \ 2)$

\quad \quad \quad $\eelse $

\quad \quad \quad \quad $ \ \sputeff \ (3\,x + 1)$

\quad \quad $\}\}$

\quad $\geteff.$

\caption{\label{fig:syracuse} Single-loop \textit{Syracuse} function.}   
\end{minipage}%
\begin{minipage}{.5\textwidth}

\noindent$\mathit{syracuse\_double}(n:\mathbb{N}) :=$

\quad $\sputeff \ n;$

\quad $ \while \ (\mdo \ x \leftarrow 
\geteff;\; \ret \ (x \not = 1) )$ 

\quad \quad $\{\{$

\quad \quad \quad $\mdo \ x \leftarrow \geteff;$

\quad \quad \quad $\iif \  \even \ x  \ \tthen$

\quad \quad \quad \quad $\while \ (\mdo \ y \leftarrow 
\geteff;\; \ret \ (\even \ y)$

\quad \quad \quad \quad  \quad  $\{\{$

\quad  \quad  \quad  \quad  \quad  \quad  $ \mdo x \leftarrow \geteff; \;$

\quad  \quad  \quad  \quad  \quad  \quad $\sputeff \ (x \ \mathit{div} \ 2)$

\quad  \quad  \quad  \quad  \quad  $\}\}$

\quad  \quad  \quad  $\eelse \ \sputeff \ (3\,x + 1)$

\quad  \quad $\}\}$

\quad  $\geteff.$
\caption{\label{fig:syracuse2} Double-loop \textit{Syracuse} function.}  
\end{minipage}
\end{figure}

The two programs are  shown in Figures~\ref{fig:syracuse} and~\ref{fig:syracuse2}. Both start with an initial natural-number value $n$, and both have a state containing a natural number: the current value in the Syracuse sequence. Both programs use while-loops, which are syntactical sugar for fixpoints defined using Kleene's theorem: $\mathit{syracuse\_single}$ (Figure~\ref{fig:syracuse}) uses one loop, and $\mathit{syracuse\_double}$ (Figure~\ref{fig:syracuse2}) uses two nested ones. 

The program $\mathit{syracuse\_double}$ can be seen as an optimized version of $\mathit{syracuse\_single}$: while the current value in the state is even, it is repeatedly divided by 2 until it is not even anymore (or forever, if the current value is 0).
In our Rocq implementation we use a binary representation of natural numbers, where repeated division by 2 amounts to removing all zeroes at the end of the number.

 Both programs use the basic primitives $\sputeff$ to write in the state  and $\geteff$ to read from it. Being monadic programs, the ``local variables" occurring in ``assignments", such as $x$ in $\mdo x \leftarrow do \ldots$, have a rather narrow ``scope", which is why reading/writing  the state occurs more often than it would in an actual imperative language with broader scope for local variables.
Everything happens as if the scope of local variables were limited to the sub-programs for the loop's condition and body, respectively.

Note also that both programs end with $\geteff$, which, given the conditions of the while-loops,  implies that they return $1$ if they terminate. The programs definitely do \emph{not} terminate when starting with initial value $n = 0$: for $\mathit{syracuse\_single}$ the (unique) while-loop does not terminate because its condition 
(state value = $1$) is always false; for $\mathit{syracuse\_double}$ it is actually the inner loop that does not terminate. Of course, the termination status for all other inputs is, in general, unknown, and, by proving their equivalence, we prove that they either both terminate (with value $1$) or do not terminate.

The relation between the two programs is formally stated as follows.

\begin{theorem}
\label{th:syracuse}
For all $n \in \mathbb{N}, \mathit{syracuse\_single} \ n \equiv_{\mathit State_{\mathbb{N}}} \mathit{syracuse\_double} \ n$.
\end{theorem}
We now give an outline of the proof. For $k \in \mathbb{N}$, let $\nbzeroes :  \mathbb{N} \to  \mathbb{N}$ be defined as follows: \linebreak
\mbox{$\nbzeroes \ 0 = \ 0$}, and, for $k > 0$, \mbox{$\nbzeroes \ k$} is the highest power of 2 that divides $k$. Next, let $\whilefuel$ be the function  recursively defined on $i\in \mathbb{N}$ by 
\medskip

\ \  \ \ \ \ \  \mbox{$\whilefuel \ 0 \ cond \ body = \bot$}
$$
\whilefuel \ (i+1) \ cond \ body = \mdo{b \leftarrow cond;
\begin{cases}
body; \whilefuel \ i \ cond \ body\  & \text{if } b = \true \\
\ret \ \ttt & \text{if } b = \false
\end{cases}}
$$
 That is, $\whilefuel  \ i \ cond \ body$ is the $i$-th iteration of a functional for the while-loop with given condition and body. By Kleene's theorem,
 $\while \ cond \ body = \lub\,\{ \whilefuel \ i \ cond \ body \mid i \in \mathbb{N}\}$.
 
  We now focus on the inner while-loop of $\mathit{syracuse\_double}$, and prove by induction on $x \geq 1$ (the value of the state before the loop) that the $\lub$ is reached in $\nbzeroes \ x$ iterations. In the remaining case $x = 0$, the $\lub$ equals $\bot$, i.e., the inner loop diverges; otherwise the loop terminates
in $\nbzeroes \ x$ iterations, and upon termination the state's  updated value  is $x / (2^{\nbzeroes \ x})$.

Next, consider the two other while-loops in the programs and their representation as $\lub$s of  while-loops with fuel given by Kleene's theorem.
We  distinguish the cases $n = 0$ and $n > 0$.

\begin{itemize}

\item if $n = 0$: we are in the case where the inner loop 
of $\mathit{syracuse\_double}$ does not terminate; its value  $\bot$ propagates to the outer loop and to the whole program; hence $\mathit{syracuse\_double} \ 0 = \bot $.
The same kind of reasoning establishes  $\mathit{syracuse\_single} \ 0 = \bot $, and 
$\mathit{syracuse\_single} \ 0 \equiv_{\mathit State_{\mathbb{N}}} \mathit{syracuse\_double} \ 0$
results from $\equiv_{\mathit State_{\mathbb{N}}}$ being reflexive, which proves our theorem for $n=0$.

 \item If $n \not = 0$, we note that, thanks to additional iterations of the inner loop, every iteration of the outer loop of $\mathit{syracuse\_double}$ corresponds to one or more iterations of the loop of $\mathit{syracuse\_simple}$, where ``corresponds to" means ``results in the same value
of the state". 
More precisely, we prove by induction on $k$ that the $k$-th iteration of the outer loop of $\mathit{syracuse\_double}$ corresponds to the iteration number $\iter \ n \ k$ of the loop of $\mathit{syracuse\_simple}$, where $\iter \ n \ k$ is the function recursively defined on $k$ shown in Figure~\ref{fig:nbiter}. Thanks to the congruence of $\equiv_{\mathit State_{\mathbb{N}}}$ w.r.t.\ $\bind$ and to properties of $\lub$ (specifically, that $\bind$ and $\lub$ commute, due to the continuity of $\bind$ in both arguments), the statement of the theorem in the current case $n \not = 0$ can be rewritten as follows:
$$\lub\,\{\mathit{syracuse\_single\_fuel} \ k \ n \mid k \in \mathbb{N}\} 
\equiv_{\mathit State_{\mathbb{N}}} 
lub\,\{\mathit{syracuse\_double\_fuel} \ (\iter \  n \ k) \ n   \mid k \in \mathbb{N}\} $$
where $\mathit{syracuse\_single\_fuel} \ k \ n$ is like $\mathit{syracuse\_single} \ n$, except that its  while loop is replaced by a while-fuel with fuel $k$; and 
$\mathit{syracuse\_double\_fuel} \ (\iter \  n \ k) \ n$ is  like $\mathit{syracuse\_double } \ n$, except that its outer while loop is replaced by a while-fuel with fuel $\iter \  n \ k$. Next, since $\equiv_{\mathit State_{\mathbb{N}}} $ is a congruence 
for $\lub$, the statement of the  theorem amounts to: for all $k,n \in\mathbb{N}$ with $n \not = 0$, $\mathit{syracuse\_single\_fuel} \ k \ n \equiv_\mathit{State_{\mathbb{N}}} \mathit{syracuse\_double\_fuel} \ (\iter \  n \ k) \ n $, which  is proved by induction on~$k$. We note that  equational reasoning with respect to the  base relation occurs in the goal of the inductive step,  simplifying and preparing it for the application of the induction hypothesis.

 \begin{figure}[ht]
\noindent \quad \quad \quad \quad $\iter \ n \ k $ :=

  \quad \quad \quad \quad \quad $\ccase \ k \ \oof$
    
   \quad \quad \quad \quad \quad \quad $0 \Rightarrow 0$

  \quad \quad \quad \quad  \quad \quad $ 1 + j \Rightarrow \iif \ n = 1 \ \tthen \ k$

  \quad \quad  \quad \quad \quad \quad \quad \quad \quad \ \  $\eelse \ \iif \ \even \ n  \ \tthen \ \iter \ (n/2^{\nbzeroes \ n}) \ j + \nbzeroes \ n$

 \quad \quad   \quad \quad \quad \quad \quad \quad \quad  \ \ $\eelse \ \iter \ (1+ 3n) \ j $

\quad \quad \quad \quad \quad $\eend$
\caption{\label{fig:nbiter} Correspondence between iterations of outer loops, when $n \not =  0$.}     
\end{figure}

\end{itemize}

\begin{remark}
The creative choices in our congruence proof regard  iterations of the loops, "counted"  using the functions $\nbzeroes$ and $\iter$. These are specific to this particular problem. The rest of the proof is specific to our approach. It uses properties of $\lub$,
in particular, that $\bind$ and $\lub$ commute, thanks to the continuity of $\bind$ in both arguments and the congruence properties  for $\bind$
and $\lub$.

\end{remark}
\section{Conclusions, Additional Related Work, and Future Work}
\label{sec:concl}
In this paper we have presented a novel domain-theoretic formalization of interaction trees in Rocq that provides a principled foundation for reasoning about effectful, potentially non-terminating computations.  
We have also demonstrated the practical utility of our approach with a non-trivial example: the equivalence  for two implementations of the possibly nonterminating Syracuse sequence.

By grounding interaction trees in domain theory, we eliminate the need for silent transitions and weak bisimulation that complicate reasoning in the closest related work~\cite{itree19}.

Another closely related work is~\emph{FreeSpec}~\cite{Letan20}, based on freer monads, where only finite program executions are considered.

\emph{Monae}~\cite{ANS19} is also limited to finite executions, and uses a  type-class-based abstraction, for which freer monads constitute a uniform mathematical representation of effectful components.

Our work is more distantly related to the algebraic approach to computational effects developed
by Hyland, Plotkin, and Power~\cite{PP01,HPP02,HPP06}. 
%They formalized effects as enriched Lawvere theories encoding operations and their equations.  
Their framework elegantly explains monad transformers as derived consequences of
effect combination, and establishes adequacy results linking operational behavior with denotational
semantics. We extend this perspective by providing a domain-theoretic formalization in the Rocq prover: where their work offers a semantic and categorical account, we deliver
a representation of effectful computation through interaction trees.

The program-equivalence part of this paper is part of substantial and growing body of literature on program equivalence for effectful languages.
Contextual equivalence for polymorphic languages with recursion and generic effects is studied by \cite{JSV10} who characterized it using logical relations.
In \cite{LGL17}, the authors develop an abstract account of applicative bisimilarity for an untyped lambda calculus with generic algebraic effects and prove that it is contained in contextual equivalence.
In \cite{SV18}, a simply-typed language with recursion and generic algebraic effects is presented along with a modal logic for expressing program behavior. This logic induces a notion of logical equivalence that is sound with respect to contextual equivalence.

Several promising directions emerge for future research.
The first one is to 
combine effects via a \emph{coproduct} operation: known to be applicable to freer monads~\cite{Swierstra08},  it can  be extended to interaction trees. This would enable us to reason about the equivalence of programs with combined effects.

The equational nature of our framework suggests opportunities for enhanced automation. Specialized tactics for effect-specific reasoning and coproduct manipulation could be developed, building on Rocq's existing rewriting capabilities to provide more seamless verification experiences.

Combining the domain-theoretic interaction trees with program logics (such as Hoare logic or separation logic) could enable even more powerful verification strategies. This integration would support mixed equational and logical reasoning about effectful programs.

% Our work establishes a solid foundation for modular, equational reasoning about effectful computations in proof assistants, bridging the gap between categorical abstractions and practical verification needs. This approach scales to increasingly complex effect interactions while maintaining the rigor and automation essential for verified software development.

%\bibliographystyle{eptcs}
%\bibliography{bibliography}

% %\appendix
% \section{Appendix.}
% We give an overview of the definition of interaction trees from the companion Rocq development.
%  The properties of completion ensure that, by hiding the details of how  $\itree \ F \ X \triangleq {\cal L} \ F \ X$ is built, interaction trees behave like  the coinductive objects in Definition~\ref{def:itree}.

 \end{document}